\documentstyle[aps]{revtex}
\begin{document}
\draft
\preprint{\vbox{\hbox{DOE/ER/40762-198}\hbox{UMD PP\#00-035}}}
\title{Crater Property in Two-Particle Bound States: When and Why}
\author{Chi--Keung Chow}
\address{Department of Physics, University of Maryland, College Park, 
20742-4111.}
\date{\today}
\maketitle
\begin{abstract} 
Crater has shown that, for two particles (with masses $m_1$ and $m_2$) in 
a Coulombic bound state, the charge distribution is equal to the sum of the 
two charge distributions obtained by taking $m_1\to\infty$ and $m_2\to\infty$ 
respectively, while keeping the same Coulombic potential.
We provide a simple scaling criterion to determine whether an arbitrary 
Hamiltonian possesses this property.  
In particular we show that, for a Coulombic system, fine structure corrections 
preserve this Crater property while two-particle relativistic corrections 
and/or hyperfine corrections may destroy it.  
\end{abstract}
Recently, in an interesting paper in this journal \cite{C}, Crater discussed 
an unusual feature of charge densities for two-particle Coulombic bound 
states.  
Let $\rho(R;m_1,m_2)$ be the charge density of a two-particle bound state 
in a given potential $V(r)\equiv V(r_1-r_2)$ in the center-of-mass coordinate 
system.  
Then Crater observed that, for a Coulombic potential the charge density 
satisfies the following relation: 
\begin{equation}
\rho(R;m_1,m_2) = \lim_{m_2\to\infty} \rho(R;m_1,m_2) + \lim_{m_1\to\infty} 
\rho(R;m_1,m_2), 
\label{C}
\end{equation}
or, in Crater's own words, one can picture $\rho(R;m_1,m_2)$ ``as equivalent 
to that produced by a particle of mass $m_1$ and charge $e_1$, bound to a 
{\it fixed\/} center with charge $e_2$ plus that produced by a particle of 
mass $m_2$ and charge $e_2$, bound to a {\it fixed\/} center with charge 
$e_1$,'' with the fixed center being the center of mass.  
Here and after this property will be referred to as the Crater property.  
Crater has shown in Ref.~\cite{C} that the Crater property holds for 
Coulombic potentials but not for generic potentials.  
In the real world, however, Coulombic potentials are often corrected by 
perturbations like fine/hyperfine structures and relativistic effects.  
It would be of interest to know what kind of corrections can be added to 
the Coulombic potential without destroying the Crater property.  
More generally, we would like to have a criterion to determine whether a 
given potential has the Crater property without explicitly solving the 
Schr\"odinger equation.  
The purpose of this paper is to provide answers to these questions.  

Let us consider an eigenfunction $\psi(r;m_1,m_2)$ to the Schr\"odinger 
equation (in units $\hbar=1$): 
\begin{equation}
H(m_1,m_2)\psi(r;m_1,m_2) \equiv \left[{-1\over 2\mu} {\partial^2\over\partial 
r_i^2} + V(r;m_1,m_2)\right] \psi(r;m_1,m_2) = E(m_1,m_2) \psi(r;m_1,m_2), 
\end{equation}
with $\mu = m_1m_2/M$ and $M = m_1+m_2$.  
The charge density operator $\hat\rho(R)$ is defined as \cite{C}
\begin{equation}
\hat\rho(R) = e_1 \delta^3(R-r_1) + e_2 \delta^3(R-r_2), 
\end{equation}
and the charge density $\rho(R;m_1,m_2)$ is its expectation value, which can 
easily be shown to be 
\begin{eqnarray}
\rho(R;m_1,m_2) &=& \int d^3r |\psi(r;m_1,m_2)|^2 \hat\rho(R) \nonumber\\
&=& e_1\left({M\over m_2}\right)^3 
\left|\psi\left({M\over m_2}R;m_1,m_2\right)\right|^2 
+ e_2\left({M\over m_1}\right)^3 
\left|\psi\left({M\over m_1}R;m_1,m_2\right)\right|^2 \nonumber\\
&=& e_1\left({m_1\over \mu}\right)^3 
\left|\psi\left({m_1\over \mu}R;m_1,m_2\right)\right|^2 
+ e_2\left({m_2\over \mu}\right)^3 
\left|\psi\left({m_2\over \mu}R;m_1,m_2\right)\right|^2 . 
\label{rho}
\end{eqnarray}

Since the eigenfunction $\psi$, satisfying the normalization condition 
$\int d^3r |\psi|^2 = 1$, carries scaling dimension [length]$^{-3/2} =$ 
[momentum]$^{3/2} =$ [mass]$^{3/2}$ (with units $\hbar=c=1$), it is always 
possible to rewrite $\psi(r;m_1,m_2)$ as 
\begin{equation}
\psi(r;m_1,m_2) = \mu^{3/2} \tilde\psi(r;m_1,m_2), 
\end{equation}
where the rescaled eigenfunction $\tilde\psi(r;m_1,m_2)$ is a dimensionless 
function.  
Now consider the case when $\tilde\psi(r;m_1,m_2)$ has the following form: 
\begin{equation}
\tilde\psi(r;m_1,m_2) \equiv \tilde\psi(\mu r), 
\end{equation}
which states that the dependences of $\tilde\psi$ on the location $r$ and the 
masses $m_{1,2}$ always come through the combination $\mu r = m_1m_2 
r/(m_1+m_2)$.  
Then $\psi(r;m_1,m_2) = \mu^{3/2} \tilde\psi(\mu r)$ and the charge density 
$\rho(R;m_1,m_2)$ in Eq.~(\ref{rho}) becomes 
\begin{eqnarray}
\rho(R;m_1,m_2)&=& e_1\left({m_1\over \mu}\right)^3 
\left|\mu^{3/2} \tilde\psi\left({m_1\over \mu}\mu R\right)\right|^2 
+ e_2\left({m_2\over \mu}\right)^3 
\left|\mu^{3/2} \tilde\psi\left({m_2\over \mu}\mu R\right)\right|^2 \nonumber\\
&=& e_1 m_1^3 \left|\tilde\psi\left(m_1 R\right)\right|^2 
+ e_2 m_2^3 \left|\tilde\psi\left(m_2 R\right)\right|^2 \nonumber\\
&=&\lim_{m_2\to\infty} \rho(R;m_1,m_2) + \lim_{m_1\to\infty} \rho(R;m_1,m_2), 
\end{eqnarray}
which is exactly the expression for the Crater property.  
In other words, the charge density exhibits the Crater property whenever the 
eigenfunction can be written as $\mu^{3/2} \tilde\psi(\mu r)$.  

It is easy to translate the above scaling condition on the eigenfunction to 
a corresponding scaling condition on the Hamiltonian.  
Since the Hamiltonian $H(m_1,m_2)$ carries scaling dimension [mass]$^1$, 
it can always be rewritten as $\mu \tilde H(m_1,m_2)$, where $\tilde H$ is a 
dimensionless function of $m_{1,2}$, as well as the relative coordinates $r$ 
and the canonical momenta $-i\partial/\partial r$.  
It is straightforward to see that $\tilde\psi$ is a function of solely $\mu r$ 
if and only if 
\begin{equation}
\tilde H(m_1,m_2) = \tilde {\cal H}\left(\mu r,
-i{\partial\over\partial (\mu r)}
\right) + \tilde V_0,
\end{equation}
such that, up to an additive constant, the masses enter the Hamiltonian only 
through combinations $\mu r$ and $-i(\partial/\partial (\mu r))$.  
The dimensionless constant $\tilde V_0$, which may have arbitrary dependences 
on $m_{1,2}$, may shift the eigenvalues but does not affect the 
eigenfunctions.  

We have shown that the charge density of an eigenfunction exhibits the 
Crater property if and only if the Hamiltonian can be written as 
\begin{equation}
H = \mu\left[\tilde {\cal H}\left(\mu r, 
-i{\partial\over\partial (\mu r)}\right) + \tilde V_0\right],  
\label{crit}
\end{equation}
which will be referred to as the scaling criterion.  
With this criterion one can easily determine if a particular potential  
exhibits the Crater property.  
For spinless Schr\"odinger systems, the kinetic term always satisfies the 
scaling criterion.  
\begin{equation}
{-1\over2\mu}{\partial^2\over\partial r_i^2} = \mu \;{-1\over2}{\partial^2
\over\partial (\mu r_i)^2}.  
\end{equation}
On the other hand, for analytic $V(r)$'s one can expand them in Laurent series 
and the scaling criterion is satisfied if and only if 
\begin{equation}
V(r) = \sum_{k=-\infty}^{+\infty} a_k \mu^{k+1} r^k + \tilde V_0, 
\end{equation}
where $a_k$ are mass independent coefficients.  
Of special interest is the case where the only non-vanishing $a_k$ are those 
with $k=-1$ and $-2$:  
\begin{equation}
V(r) = {a_{-1}\over r} + {a_{-2}\over \mu r^2}, 
\end{equation}
which describes a Coulombic potential in three dimensions with $a_{-1} = 
e_1e_2$ and $a_{-2} = \ell(\ell+1)/2$, {\it i.e.}, the case studied in 
Ref.~\cite{C}. 
Another interesting case is the ``two-dimensional Coulombic potential'', 
{\it i.e.}, the logarithmic potential:  
\begin{equation}
V(r) = e_1e_2 \ln(r/r_0) = e_1e_2 \left[\ln(\mu r) - \ln(\mu r_0)\right],  
\end{equation}
which can be seen to satisfy the scaling criterion by identifying the second 
term as $\tilde V_0$.  

As pointed out in Ref.~\cite{C}, the Crater property is {\it not\/} a feature 
of potentials of generic $r$ and mass dependences. 
Crater illustrated this point by studying the eigenfunctions of a simple 
harmonic potential and showed explicitly that $\rho(R;m_1,m_2)$, given 
by Eq.~(\ref{rho}), is {\it not\/} the sum of $\lim_{m_2\to\infty} 
\rho(R;m_1,m_2)$ and $\lim_{m_1\to\infty} \rho(R;m_1,m_2)$, {\it i.e.}, the 
simple harmonic potential does not exhibit the Crater property.  
On the other hand, we can reproduce the same conclusion by just studying 
the scaling behavior of the simple harmonic potential: 
\begin{equation}
V(r) = {1\over2} \mu \, \omega^2 r^2, 
\end{equation}
which cannot be recast in a form conforming to criterion (\ref{crit}).  
As a result, the Crater property is not exhibited in simple harmonic 
potentials.  
 
It is of interest to note that, for any potential $V(r) = \mu \tilde V(\mu r)$ 
satisfying the scaling criterion, including the fine structure corrections 
does not destroy the Crater property.  
\begin{eqnarray}
H_{mv} &=& {-1\over 8\mu^3 c^2} \left({\partial^2\over\partial r_i^2}\right)^2 
= \mu\;{-1\over 8 c^2}\left({\partial^2\over\partial (\mu r_i)^2}\right)^2, \\
H_{SO} &=& {1 \over \mu^2 c^2} {1\over r} {dV(r)\over dr} L\cdot S = 
\mu\;{1 \over c^2} {1\over \mu r} {d\tilde V(\mu r)\over d(\mu r)} L\cdot S, \\
H_{D} &=& {1\over 8\mu^2 c^2} {d^2 V(r)\over dr_i^2} = \mu \;{1\over 8 c^2} 
{d^2 \tilde V(r)\over d (\mu r_i)^2},  
\end{eqnarray}
where $H_{mv}$, $H_{SO}$, and $H_D$ stand for the relativistic mass variation 
term, the spin-orbit coupling term and the Darwin term, respectively.  
This may look miraculous but is actually nothing but a consequence of the 
fact that all these fine structure corrections come from the leading order 
expansion in the fine structure constant of the one-particle Dirac Hamiltonian 
with the same potential $V(r)$.  
\begin{equation}
H = -i \alpha \cdot {\partial\over\partial r} + \beta \mu + V(r) = 
\mu\; \left[-i \alpha \cdot {\partial\over\partial (\mu r)} + \beta + 
\tilde V(\mu r)\right],  
\end{equation}
where $\alpha$ and $\beta$ are the Dirac matrices.  
Since this one-particle Dirac Hamiltonian also satisfies the scaling 
criterion, the Crater property is preserved.  

However, it is important to bear in mind that it is an approximation to 
describe a two-particle bound state by a one-particle Schr\"odinger or 
Dirac equation.  
Take, for example, the hyperfine correction, which for a Coulombic bound 
state is 
\begin{equation}
H_{hf} = {g_1g_2 e_1e_2\over 3 m_1m_2} S_1\cdot S_2 \, \delta^3(r) 
= \mu \left[\mu\over M\right] {g_1g_2 e_1e_2\over 3} S_1\cdot S_2 \, 
\delta^3(\mu r), 
\end{equation}
and the scaling criterion is violated by the outstanding factor of $\mu/M$, 
where $M=m_1+m_2$ is the total mass.  
Violations of the scaling criterion may also be due to two-particle 
relativistic effects.  
In the non-relativistic theory a two-particle problem can always be reduced 
to an effective one-particle problem in the relative coordinates, which  
decouple with the center-of-mass coordinates in the kinetic term:
\begin{equation}
{1\over 2m_1}{\partial^2\over\partial r_1^2} + 
{1\over 2m_2}{\partial^2\over\partial r_2^2} =
{1\over 2M}{\partial^2\over\partial R^2} +
{1\over 2\mu}{\partial^2\over\partial r^2}, 
\end{equation}
where $R$ is the center of mass position $m_1r_1+m_2r_2$, and $r$ is the 
relative position $r_1-r_2$. 
For a relativistic theory in general no such decomposition is possible, 
and the description of a two-particle problem by a one-particle equation is a 
good approximation only when one particle is much heavier than the other.  
Such treatments do not capture genuine two-particle effects, like the 
two-particle relativistic corrections and hyperfine corrections.  
These corrections in general do not satisfy the scaling criterion and 
one expects the Crater property to be violated by these corrections. 

In passing, we note that the notion of Crater property can be generalized in a 
straightforward manner to any operator of the following form: 
\begin{equation}
{\cal O}(R) = a \, \delta^3(R-r_1) + b \, \delta^3(R-r_2), 
\end{equation}
where $a$ and $b$ are arbitrary coefficients.  
This operator ${\cal}(R)$ may correspond to physically interesting objects 
for particular choices of $a$ and $b$; it is the charge density when 
$(a,b)=(e_1,e_2)$, the probability density of particle 1 and 2 when $(a,b)
=(1,0)$ and $(0,1)$, respectively, and the mass density when $(a,b)=(m_1,m_2)$.
Then the Hamiltonian or potential is said to exhibit the Crater property of 
charge/probability/mass distribution if and only if the charge/probability/mass
distribution in the bound state of particle mass $m_1$ and $m_2$ is equivalent 
to the sum of the charge/probability/mass distributions produced in the limits 
$m_1\to \infty$ and $m_2\to \infty$.  
As before, all these Crater properties are guaranteed by the same scaling 
criterion (\ref{crit}).  

In conclusion, we have provided a simple criterion to determine if the 
eigenfunctions of a given Hamiltonian have the Crater property.  
In particular, we have shown that neither the inclusions of fine structure 
corrections nor the switching from Schr\"odinger to Dirac formalism will 
destroy the Crater property.  
The author believes Crater must have foreseen the essential points of this 
paper --- as in the conclusion of Ref.~\cite{C} he stated that ``in general, 
the appearance of parameters in the potentials that are not dimensionless (in 
natural units) and do not depend on the reduced mass would not be of the 
correct type.''
As a consequence, this paper may be regarded as a concrete realization of 
this observation.  

\acknowledgements
This work is supported by the U.S.~Department of Energy grant 
DE-FG02-93ER-40762.


\begin{references}
\bibitem{C} Horace W.~Crater, {\sl ``An Unusual Feature of Charge Densities 
for Two-Particle Bound States''}, Am.~J.~Phys.~{\bf 67} 739 (1999).  
\end{references}
\end{document}